\documentclass[twocolumn,prb,amsmath,showpacs,amssymb,citeautoscript,floatfix]{revtex4}
\def\etal{{ \it et al. }}
\def\prb{{Phys. Rev. B }}
\def\pl{{Phys. Rev. Lett. }}
\usepackage{epsfig}

\begin{document}
\title{First principles theoretical studies of half-metallic ferromagnetism in CrTe
} 
\author{Y. Liu}
\affiliation{Physics Department, Brock University, St. Catharines,
Ontario L2S 3A1, CANADA                    }
\affiliation{College of
Science, Yanshan University, Qinhuangdao, Hebei 066004, China}
\author{S. K. Bose}
\affiliation{Physics Department, Brock University, St. Catharines,
Ontario L2S 3A1, CANADA                    } 
\author{J. Kudrnovsk\'{y}}
\affiliation{Institute of Physics, Academy of the Sciences of the Czech Republic, Na Slovance 2,
182 21 Prague 8, Czech Republic }
\begin{abstract}
Using  full-potential linear augmented plane wave method (FP-LAPW) and the density functional
theory, we have carried out a systematic investigation of the electronic,
magnetic, and cohesive properties of the chalcogenide CrTe in three competing  structures:
rock-salt (RS), zinc blende (ZB) and the NiAs-type (NA) hexagonal.
Although the ground state is of NA structure,  RS and ZB are interesting in that these fcc-based structures, which can possibly be grown on
many semiconductor substrates, exhibit half-metallic phases
above some critical values of the lattice parameter. We find that the NA structure is not half-metallic at its equilibrium volume, while both
ZB and RS structures are. The RS structure is more stable than the ZB, with an energy that is lower
by 0.25 eV/atom. While confirming previous results on the half-metallic phase in ZB structure, we provide hitherto
unreported results on the half-metallic RS phase, with a gap in the minority channel and a  magnetic moment
of $4.0$ $\mu_{B}$ per formula unit. A comparison of
total energies for the ferromagnetic (FM), non-magnetic (NM), and 
antiferromagnetic (AFM) configurations 
shows the lowest energy configuration to be FM for CrTe in all the three structures. The FP-LAPW calculations are supplemented by 
linear muffin-tin orbital (LMTO) calculations using both local density approximation (LDA) and LDA+U method. The exchange interactions and the Curie temperatures calculated via the linear response method in ZB and RS CrTe are compared 
over a wide range of the lattice parameter. The calculated Curie temperatures for the RS phase are consistently higher than those for the
ZB phase.
\end{abstract}
\maketitle 
\section{Introduction} 

Half-metallic (HM) ferromagnetism\cite{Pickett2001} is an ideal property of materials from the viewpoint of their application in spintronic
devices\cite{Zutic2004,Prinz-science-282-1660-1998,Wolf-science-294-1488-2001}. 
Thus a considerable research has been devoted in recent years to the
designing and measurement of the physical properties of such 
materials. One group of such materials is known as the dilute magnetic semiconductors (DMS)\cite{Saito,Sato4,Sato5},
where the magnetic properties are mediated by the percolation effects connecting a very small concentration of
transition metal (TM) atoms such as Mn or Cr in a semiconductor host. Another group, that has been studied extensively, consists of 
TM-semiconductor binary/ternary alloys, where the TM concentration can be as high as the other 
components\cite{Hong-jap-97-063911-2005,Kronik-prb-66-041203-2002,Yong-jmmm-307-245-2006}. 
In particular, some ZB structure transition metal chalcogenides and
pnictides have been predicted to be HM ferromagnets on the basis of theoretical
calculations\cite{Brener-prb-61-16582-2000,Zhao-prb-2001,Pettifor2003,Liu-prb-67-172411-2003,
Galanakis-prb-66-134428-2002,Galanakis-prb-66-012406-2002,Galanakis2003,Sanyal2003,Kurmaev-prl-84-144415-2000,Wijs-69-214429-2004,
Nakao2004,Zheng2004,Groot-rmp-80-315-2008}. Here the magnetic properties of the transition metal matrix is modified via the hybridization between the transition metal 3-$d$ and the
$s$ and $p$ orbitals of the pinctogen or the chalcogen. Half-metallicity arises as a result of interplay between the exchange splitting and the
hybridization.
 Some
Cr-based chalcogenides and pnictides\cite{Akinaga2000,Li2008,Shirai2003,Akinaga2005,Yamana2004,Kahal2007,Galanakis2003,Pask,Ito,Shi,Zhang,Kubler2003},
have been predicted to be not only HM but also potential candidates for possessing high Curie temperatures $T_c$.  Akinaga \etal \cite{Akinaga2000} have reported  growing ZB thin films of 
CrAs on GaAs (001) 
substrates via molecular beam epitaxy. These showed ferromagnetic behavior at temperatures in excess 
of 400 K and magnetic moments of 3$\mu_B$ per CrAs unit.  Thin films of CrSb grown similarly
 on GaAs, (Al,Ga)Sb, and GaSb substrates have been found to be of ZB structure and ferromagnetic with $T_c$ 
higher than 400  K\cite{Zhao2001}. Deng \etal \cite{Deng2006} were successful in increasing the 
thickness of ZB-CrSb films to $\sim$ 3 nm  using (In,Ga)As buffer layers, and Li \etal \cite{Li2008} were able to grow 
$\sim$ 4 nm thick ZB-CrSb films on NaCl (100) substrates. Two of the present authors have recently presented an extensive
theoretical study of the exchange interactions and Curie temperatures in a series of Cr-based pnictides and chalcogenides
in ZB structure as a function of the lattice parameter\cite{Bose2010}.

One important issue is that although the ZB phase of these compounds shows half-metallicity and often (the promise of) high Curie
temperature, the ground state usually has a different structure. For many of these compounds the ground state is NiAs-type hexagonal (NA).
The semiconductor substrates on which thin films of these compounds are grown have cubic structures, usually  RS or 
ZB. The other point is that often the equilibrium volume does not show half-metallicity, which appears only at an expanded volume. From this
stand-point it would be useful to explore the possibility of the half-metallic phase in structures other than ZB, and in particular the RS structure. In this work we have examined this problem for CrTe. Zhao and Zunger\cite{Zunger2005}  argued that ZB CrTe should be epitaxially
unstable against the NA structure. However, recently Sreenivasan \etal \cite{Sreenivasan2008} and Bi \etal \cite{Bi2008} have reported growing $\sim$5 nm thick thin films of ZB CrTe by low temperature molecular beam epitaxy. Their measurements of temperature-dependent
remanent-magnetization indicate a Curie temperature of $\sim$100 K in these films. One attractive feature of bulk CrTe, at least from a theoretical view point, is that the equilibrium volume in the ZB structure is already half-metallic, whereas most other Cr-based chalcogenides
and pnictides show half-metallicity only at expanded volumes\cite{Galanakis2003}. Besides, the half-metallic gap is larger in the chalcogenides like CrTe or CrSe than in the pnictides such as CrAs or CrSb. This is due to larger Cr-moment in the half-metallic phase
 for the chalcogenides ($\sim$4$\mu_B$) than for the pnictides ($\sim$3$\mu_B$), which results in larger exchange splitting. Our previous study\cite{Bose2010}
 has revealed that chalcogenides such as CrS and CrSe have a high likelihood of exhibiting antiferromagnetic or complex magnetic behavior
for a range of the lattice parameter, whereas such tendencies are less pronounced in CrTe. Hence, the focus of this work is on CrTe.
 We study the appearance of half-metallicity in three competing structures for CrTe: NA, RS and ZB.  The ground state structure (NA) shows half-metallicity not at the equilibrium volume but  at some expanded volume, while both RS and ZB structures show half-metallicity at their respective equilibrium (lowest energy) volumes.  The RS structure is found to be more stable than the ZB, with an energy that is lower by 0.25 eV/atom. 
Some band calculations for CrTe in the NA structure have already
appeared\cite{Dijkstra-jpcm-1-9141-1989} in the literature, and the same goes for ZB CrTe\cite{Galanakis2003,Pettifor2003}.  Ahmadian \etal\cite{Ahmadian2010} have recently discussed half-metallicity at a ZB CrTe(001) surface and its 
interface with ZnTe(001). So far CrTe in the RS structure has not been studied from the viewpoint of its electronic and magnetic properties. Although both RS and ZB structures are fcc-based, their magnetic properties should differ somewhat for the same lattice parameter due to different levels of hybridization between the Cr 3-$d$ and Te $sp$-orbitals. This should influence the magnetic moments, half-metallic gaps and the exchange interactions. We find that
the exchange interactions are predominantly ferromagnetic 
and strong in the RS structure, while in the ZB structure Cr-Cr interactions can have non-negligible antiferromagnetic components for some lattice parameters. This results in a lower Curie temperature $T_c$ for the ZB structure over a wide range of the lattice parameter. The promise of lower energy and higher $T_c$  should prompt the experimentalists to try to fabricate CrTe thin films in the RS structure. 

\section{Electronic structure}

The two fcc-based structures ZB ($F\bar{4}3m$, No.$216$) and RS ($Fm\bar{3}m$, No.$225$) are well-known. The NA
structure(P6$_{3}$/mmc, No.$194$) is based on a distorted hcp array of the  arsenide anions.
 By contrast with the wurtzite structure, which shares the same formula type (AB) with the NA structure, the cations  occupy all the octahedral sites rather than half the tetrahedral holes. There is one octahedral hole for each hcp lattice site, and so the AB stoichiometry is preserved. The local co-ordinations of the anions and cations are different in the NA and wurtzite structures. Bulk samples of CrTe are known to crystallize  
in this structure\cite{Chevreton1963} with cell parameters $a=3.997$\AA, $c=6.222$\AA$\;$ at room temperature. The Cr-Te bond length is 
2.78\AA$\;$ and
the $c/a$ ratio is 1.557, different from the ideal hcp value of 1.633. Thus the six Te atoms form a trigonally distorted octahedron around Cr. 
The Cr-Cr nearest neighbor distance in the $ab$ plane is $a=3.997$\AA, while the out of plane distance is $c/2=3.111$\AA. Readers may consult Fig. 1 of Ref.[\onlinecite{Dijkstra-jpcm-1-9141-1989}] to view the structure.

For the electronic structure study we have used the WIEN$2$k code \cite{Blaha-cpc-59-399-1990,Blaha-wien2k}, which employs the
full-potential linear augmented plane waves (FP-LAPW) plus local orbitals method. The generalized gradient
approximation (GGA) proposed by Perdew \etal was used for exchange
and correlation potentials\cite{Perdew-prl-77-3865-1996}. According
to a study by Continenza \etal
\cite{Continenza-prb-64-085204-2001}, GGA is essential for obtaining
accurate equilibrium volumes and magnetic moments for Mn-based pnictides of ZB and NA structures.
 We consider full
relativistic effects for the core states and use the scalar
approximation for the valence states. Spin-orbit coupling is neglected, as it should have negligible effect on the
results for CrTe.  Convergence with respect to basis set and k-point sampling
was carefully verified.
About $3000$ K points were used for the Brillouin zone (BZ) integrations, using
the  Monkhorst-Pack scheme\cite{Monkhorst-prb-13-5188-1976} and $14 \times14
\times14$ divisions of the BZ.  We used $R_{mt}*K_{max}=8.0$ and angular momentum expansion up to
$l_{max}=10$ in the muffin tins, and $G_{max}=14$ for the charge
density.  All core and valence states were treated self-consistently. For the
calculations involving the different structures the muffin-tin (MT)
radii were taken to be equal to $2.2$ and $2.5$ bohrs for Cr and Te
atoms, respectively.  The same MT radii were used in the
calculations of different structures at their equilibrium volumes in order to 
compare accurately the total energies. Self-consistency in charge density was achieved to an accuracy
higher than $10^{-4}$.

We  carried out a systematic structural optimization of NA, ZB, and RS CrTe  in FM, NM and
AFM configurations. For the NA structure the c/a ratio was optimized and the result $c/a=1.542$ for the FM phase is
in good agreement with previously obtained result\cite{Ahmadian2010}. 
This optimized c/a ratio was used for the 
final NA structure volume optimization.
Bulk modulus  was calculated by fitting the
volume vs. energy results  to the empirical
Murnaghan equation of state\cite{Murnaghan-pnas-30-244-1944}. We find that for all three structures the FM phase has
lower energy than the NM  and AFM phases.  To assess the relative stability of CrTe in different structures we
consider the total energy per formula unit in the ZB and RS
structures relative to the NA structure. The
calculated optimized equilibrium lattice constant a, bulk modulus B, the total energy difference E$_{1}$ between the NM and the FM
states: $E_{1}=E_{NM}-E_{FM}$,  and E$_{2}$ between the AFM and the FM
states: $E_{2}=E_{AFM}-E_{FM}$,  and the metastability energy $E_{m}$, which is the total energy per formula
unit relative to the NA structure, are given in Table\ref{table1}. Note that for RS and ZB structures, in calculating $E_{2}$ we have considered the energies of three AFM configurations:
[001],[110] and [111] and considered the lowest of these energies. For the hexagonal NA structure, the AFM state was created by simply ascribing
equal but opposite magnetic moments to the two Cr atoms in the unit cell. Cohesive properties are dependent on the treatment of the exchange-correlation effects in a non-trivial manner, as indicated by the LDA values of the lattice parameter and bulk modulus shown in parentheses
in  Table\ref{table1} for the RS and ZB equilibrium structures, where all other values are from GGA. Increased correlation effects in GGA increases
the equilibrium lattice parameter and lowers the bulk modulus.
\begin{table*}
\label{table1}
\caption{Results obtained via the WIEK2$k$ code: equilibrium lattice constant a, Cr-Te
bond-length L$_{CrTe}$, cohesive energy $E_c$ and bulk modulus B 
for ZB, RS and NA CrTe in the FM state. $E_c$ is defined as negative of the total energy of the formula unit with respect to the sum
of the energies of the free Cr and Te atoms. E$_{1}$, E$_{2}$ and E$_{m}$ are as defined in the text. All results shown are obtained using GGA.
In some cases the corresponding LDA values are shown in parentheses.}
\begin{ruledtabular}
\begin{tabular}{lccccccc}
Structure &  a (\AA)        & L$_{CrTe}$(\AA) & B(GPa)&E$_c$(eV)     &E$_{1}$(eV)& E$_{2}$(eV)& E$_{m}$(eV)  \\
\hline
ZB& 6.263 (6.088)      & 2.712      & 45.278 (57.962) &6.13 & 4.416 & 0.284  & 0.328\\
RS& 5.727 (5.533)      & 2.864      & 59.744 (70.823)&6.38  & 1.605 & 0.019  & 0.076\\
NA (c/a=1.542) & 4.118      & 2.859   & 46.487 & 6.46 & 1.453 & 0.107 & 0 \\
\end{tabular}
\end{ruledtabular}
\end{table*}

\begin{figure}
\includegraphics[angle=270,width=3.5in]{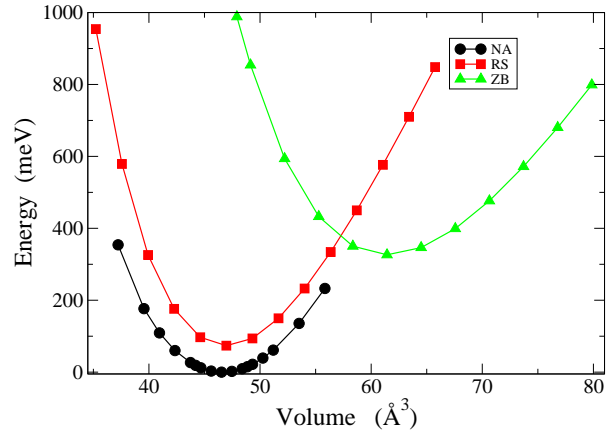}
\caption[]{(Color online) Energy as a function of volume per atom in the three structural phases of CrTe.  }
\label{fig1}
\end{figure}

It is apparent from Table\ref{table1} that the FM state is lower in energy than the NM and AFM states for all
the three structures and the ground state structure is NA.
Our calculated CrTe results of ZB and NA by GGA are in good agreement with previous results\cite{Pettifor2003,Ahmadian2010}. 
The calculated total energy as a function of the atomic volume for FM phase ZB, RS, and
NA structures of CrTe is shown in Fig.\ref{fig1}.   The ZB and RS phases are higher in
energies than NA by $0.328$ eV and $0.076$ eV/atom, respectively. However, the RS structure is only marginally unstable against
the NA structure, and more stable than the ZB structure, with an energy that is about 0.25 eV/atom lower. Thus we have carried out a detailed
study  of the RS phase.
For  lack of both experimental and
theoretical data regarding RS CrTe, the present
results should be considered as predictions and are intended to serve as reference for future experimental work.

As the density of states (DOS) for the ZB structure has been described in several 
publications\cite{Pettifor2003,Galanakis2003,Shi,Zhang}, we present only the DOS's for the RS and NA structures in this work. For the RS structure
the majority-spin channel, as in case of the ZB structure, is metallic, while the minority-spin states have an energy gap of width 0.76 eV around the Fermi level. The minority-spin gap for the ZB structure is 2.9 eV. All of these values refer to the respective equilibrium volumes.  Thus both
ZB and RS phases are HM systems. The HM gap, which is defined in Ref.[\onlinecite{Pettifor2003}] is 0.83 eV for the equilibrium ZB volume, while
for the RS equilibrium volume it is very small, being about 0.03 eV. However, it increases with increasing volume. The spin-polarization at the
Fermi level is quite small  for RS CrTe, but the half-metallicity is robust against volume expansion.
 The general structure of DOS is similar around the Fermi level for both
structures. A closer inspection, however, reveals some
differences. 
In the spin-up channel there is a gap in the energy interval between $1.62$ and $2.82$ eV just below the Fermi level for the ZB structure. On the other hand, for the spin-down
DOS the gap at the Fermi level in  ZB CrTe is larger than that in RS CrTe. Also the peak structures differ by
an energy shift with respect to the Fermi level. All these differences should be accountable on the basis of different levels of hybridization between the Cr-3$d$ and Te-5$p$ orbitals and different exchange splittings for the two structures. The growth of metastable ZB-CrAs\cite{Akinaga2000}, ZB-CrSb\cite{Zhao2001,Deng2006} and ZB-CrTe\cite{Sreenivasan2008,Bi2008} indicates the suitability of RS-CrTe for growth on carefully chosen substrates. The equilibrium lattice constant of RS CrTe (5.727 \AA) is only 4.7\% smaller than the equilibrium lattice parameter 6.013 \AA$\;$ of RS GeTe (both values quoted are from calculations). Thus RS GeTe should act as a suitable substrate for growing RS CrTe films. 

\begin{figure}
\includegraphics[angle=270,width=3.5in]{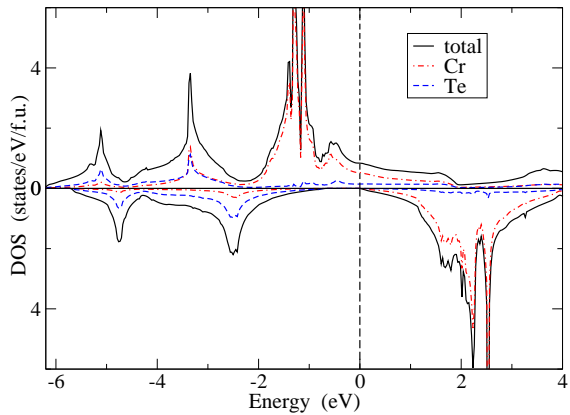}
\caption[]{(Color online)  DOS for the RS structure at the equilibrium volume (a=5.7274 \AA) obtained via the FP-LAPW method. }
\label{fig2}
\end{figure}

\begin{figure}
\includegraphics[angle=270,width=3.5in]{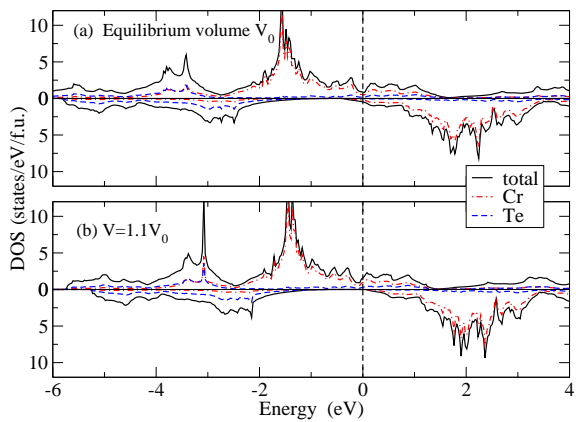}
\caption[]{(Color online)  DOS for the NiAs structure at the (a) equilibrium volume V$_0$ (c/a=1.5424, a=4.118 \AA) and
(b) 10\% expanded volume V=1.1V$_0$ (c/a=1.5424), obtained via the FP-LAPW method. }
\label{fig3}
\end{figure}

In Figs.\ref{fig2} and \ref{fig3} we show the DOSs for the RS and NA structure CrTe at their respective equilibrium volumes obtained via 
WIEN2k FP-LAPW code. Note that at equilibrium volume CrTe in NA structure is not half-metallic, but becomes so at expanded volumes.
\begin{figure}
\includegraphics[angle=270,width=3.75in]{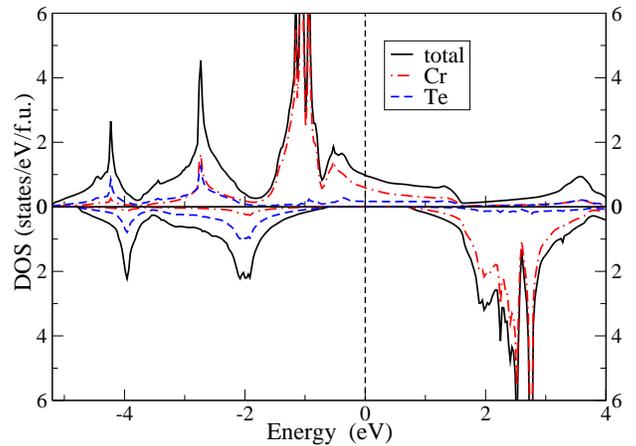}
\caption[]{DOS in RS CrTe at the equilibrium lattice
parameter of  RS-GeTe (6.013 \AA).} 
\label{fig4}
\end{figure}
 In Fig.\ref{fig4}, the DOS of RS CrTe at the equilibrium RS-GeTe lattice parameter (6.013 \AA) is given, where it is seen to be half-metallic. The minority-spin gap is $1.388$ eV, HM gap is $0.637$ eV. Both the minority-spin gap and HM gap become larger with increasing lattice 
parameter and half-metallicity is robust with
respect to lattice expansion. 

\begin{figure}
\includegraphics[angle=270,width=3.5in]{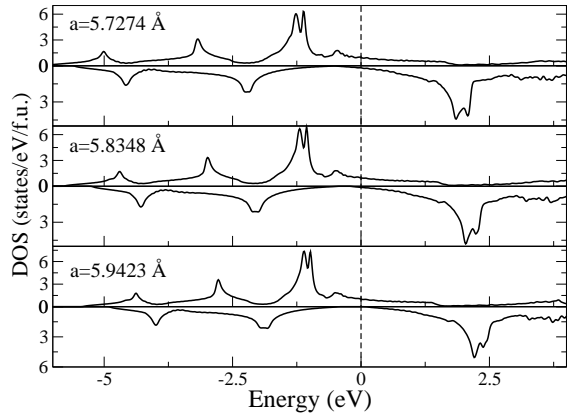}
\caption[]{ LMTO-ASA DOS for the RS structure for three different lattice parameters, from just below to just above the critical value for
the HM phase.}
\label{fig5}
\end{figure}

We have supplemented the FP-LAPW calculations with LMTO-ASA calculations using the tight-binding linear muffin tin orbital (TB-LMTO)\cite{Andersen} 
  method and its implementation for random alloys using the coherent potential approximation (CPA)\cite{Kudrnovsky1990,Turek97}.  The exchange-correlation potential 
given by Vosko, Wilk and Nusair\cite{VWN} is used.  We optimize the ASA (atomic sphere 
approximation) errors by including empty spheres in the unit cell. 
The DOS for the RS structure obtained via the LMTO-ASA method for three different lattice parameters, from just below to just above the critical value for
the half-metallic phase is shown in Fig.\ref{fig5}.

As pointed out above, several features of the electronic structure are dependent on the exchange-correlation effects. Enhanced correlation 
generally increases the equilibrium lattice parameter and the magnetic moment for a given volume. The equilibrium lattice parameter for the ZB structure is 6.26 \AA$\;$ according to the FP-LAPW calculations using GGA, 6.09 \AA$\;$ according to FP-LAPW and LDA, and 6.03 \AA$\;$ 
according LMTO-ASA method using LDA. These LDA values value are in close agreement with the result 6.07 \AA$\;$ from the Korringa-Kohn-Rostoker (KKR) Green function method and LDA\cite{Galanakis2003}. For RS CrTe equilibrium lattice parameter from FP-LAPW GGA is 5.727 \AA, while replacing GGA by LDA results in the value 5.53 \AA, a decrease of about 3\%. Since GGA is not incorporated at this stage in the LMTO-ASA code that we used, we checked the effect of enhancing the correlation effect in LMTO-ASA by using the LDA+U method\cite{Anisimov1997}. The equilibrium lattice parameter for RS CrTe increases from 5.245 \AA$\;$ to 5.4\% \AA, i.e. by about 3\%,  from LDA to LDA+U with U=1.02 eV (0.075 Ry) and J=0.
Note that the width of the energy gap in the minority states for the RS structure also has a noticeable dependence on  the treatment of exchange and correlation effects. This gap is widened with enhancement of correlation effects. For example, for a lattice parameter just above the critical value
for the half-metallic state (Fig.\ref{fig5}), the LMTO-ASA calculation with LDA gives a gap width of approximately 0.5 eV, while LDA+U calculation
for U=0.075 Ry (1.02 eV), the gap increases to $\sim$0.8 eV, which compares well with the FP-LAPW GGA value of 0.76 eV.
\begin{figure}
\includegraphics[angle=270,width=3.5in]{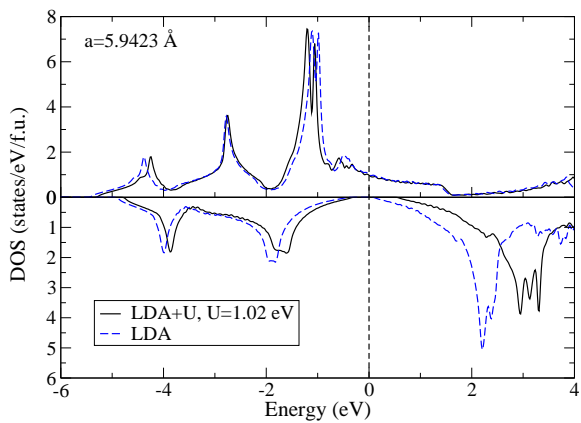}
\caption[]{ LMTO-ASA DOS for the RS structure using the LDA and LDA+U (U=1.02 eV) method for a lattice parameter of 5.9423 \AA.}
\label{fig6}
\end{figure}

\section{Magnetic moments}
\begin{figure}
\includegraphics[angle=270,width=3.5in]{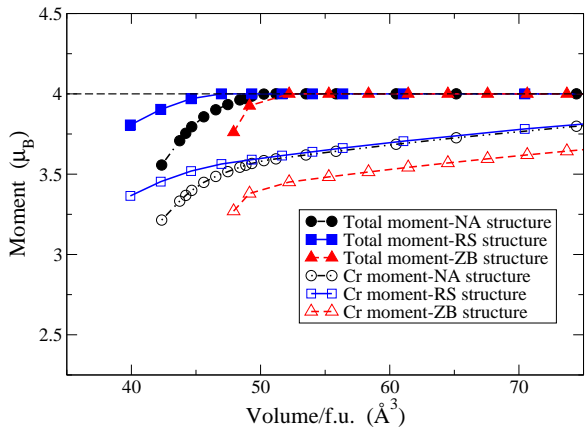}
\caption[]{(Color online) Magnetic moment per formula unit (f.u.) for the NA, ZB and RS structures obtained via FP-LAPW 
(WIEN2$k$) calculations.  }
\label{fig7}
\end{figure}

Results of the WIEN2$k$ calculations for magnetic moments and some other useful quantities are given in Table \ref{table2}.
The total magnetic moment
contains  contributions  from the Cr atom, the Te atom and the
interstitial region. The Cr atom carries the largest
amount of magnetic moment, the Te site has a small magnetic
moment, about 3-4\% of that of the Cr atom  and a certain amount of moment, 16-17\% of the Cr-moment, is also found in the
interstitial region. The magnetic moment in the interstitial region
arises mainly from the d-states of the Cr atom. The moments of Cr and Te are antiparallel to each other, while the moments of the
interstitial region and the Cr atom are parallel.
 The calculated total magnetic moment for both ZB and RS structures are 4.000 $\mu_B$ per formula unit at their equilibrium lattice parameters. 
An integer value of the magnetic moment is a
characteristic feature of HM ferromagnets. The ratio between the induced
moment on Te and the moment on Cr is $0.044$, $0.031$ and $0.030$ for ZB, RS and
NA, respectively. 
 The magnitudes of the moments on the two
sublattices increase with increasing lattice parameters, due to decreased
hybridization between Cr-$d$ and Te-$sp$ orbitals. Above a critical
value of the lattice parameter, the moment per formula unit (f.u.)
saturates at a value of 4.000 $\mu_B$, as the HM state is achieved,
while the local moments on the Cr- and Te-sublattices increase in
magnitude, remaining opposite to each other.     The saturation value
 satisfies the so-called 
``rule of 8''\cite{Galanakis2003}: $M=\left(Z_{tot}-8\right)\mu_{B}$, where $Z_{tot}$ is the total number of valence 
electrons in the unit cell. The number 8 accounts for the fact that in the HM state 
the bonding $p$ bands are full, accommodating 6 electrons and so is the low-lying band formed 
of the $s$ electrons from the $sp$ atom, accommodating 2 electrons.  The magnetic moment 
comes from the remaining electrons filling the $d$ states, first the $e_g$ states and then the 
$t_{2g}$.   The moment versus volume variations for the three structures obtained in the WIEN2$k$ calculations are shown in Fig.\ref{fig7}.

\begin{table*}

\caption{Comparison of magnetic moments per formula unit ($\mu_B$), minority-spin gap G$_{min}$ (eV) and HM gap G$_{HM}$ (eV) for FM CrTe 
in ZB, RS and NA structures, at their respective equilibrium volumes, obtained by using the FP-LAPW method (WIEN2$k$). The numbers without and
within the parentheses refer to values obtained by using GGA and LDA, respectively.}\label{table2}
\begin{ruledtabular}
\begin{tabular}{lcccccc}
 Structure  &  Cr   & Te    & Interstitial   & Total  &G$_{min}$ & G$_{HM}$\\
\hline
ZB   & 3.542 (3.501) &-0.157 (-0.119) & 0.616 (0.619) &4.000 (4.000) & 2.911   & 0.829 \\
RS   & 3.563 (3.517) &-0.112 (-0.079) & 0.548 (0.548) &4.000 (3.986) & 0.761   & 0.028\\
NA   & 3.489 (3.398) &-0.105 (-0.076) & 0.520 (0.503)  &3.904 (3.825) & -       & - \\
\end{tabular}
\end{ruledtabular}
\end{table*}

In LMTO-ASA method the interstitial region is described by empty spheres and the electronic properties of this region is primarily dictated by the
tails of the orbitals centered at the real atoms such as Cr and Te. As in the FP-LAPW calculation, we find that the magnetic moments of the empty spheres are parallel to those of the Cr atoms, while those of the Te atoms are antiparallel. For the RS structure, the moment on the Te atom is small (around 1-2\% of that of the Cr atom) for low values of the lattice parameter, while the moment on the two empty spheres are each about 3.5-4\% of that of the
Cr atom. With increasing lattice parameter, the Te-moment increases with respect to the Cr-moment to about 7-8\% at the highest lattice parameter
studied. At these high lattice parameters the moments on the empty spheres, originating from the tails of the nearby Cr-3$d$ orbitals, drop to
about 2\% of the Cr-moment. 

\begin{figure}
\includegraphics[angle=270,width=3.5in]{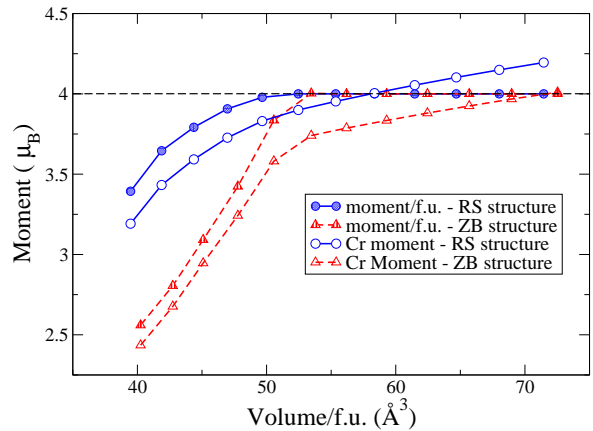}
\caption[]{(Color online) Magnetic moment per formula unit (f.u.) and of the Cr atoms in the ZB ans RS structures obtained in the
LMTO-ASA calculations.  }
\label{fig8}
\end{figure}

In the ZB structure, the two empty spheres are not equivalent and the induced moments on the empty spheres and the
Te atoms in this case has been described in detail in our recent publication\cite{Bose2010}. 
For the ZB structure the induced moment on Te is negligible for small lattice parameters, increases steadily with increasing lattice parameter, achieving a maximum value of 4\% of the Cr-moment
at the highest value of the lattice parameter studied. Most often it is less than 1\%.
 The magnitudes of the induced moments on the empty spheres
decrease in the following order: 
sublattice ES-1 (the sublattice of empty spheres that is at the same distance with respect to the Cr-sublattice
 as the Te-sublattice ),
sublattice ES-2 (sublattice of empty spheres further away from the Cr-sublattice).  
The induced moment on ES-2 is always small, decreases with increasing lattice parameter from about 2\%  to about
0.4\% of the Cr-moment. The induced moment on ES-1 is higher than the other two induced moments for smaller
lattice parameters. For larger lattice parameters it is comparable to the Te-moment, but opposite in sign.
Above a critical value of the
lattice parameter, the moment per formula unit (f.u.) saturates at a value of 4.0 $\mu_B$,
as the HM state is achieved,
while the local moments on the Cr- and Te-sublattices increase in magnitude, remaining opposite in sign.
Sandratskii 
\etal\cite{Sandratskii} have discussed the problem associated with such 'induced moments' in calculating the Curie temperature.
LMTO-ASA results using LDA for the ZB and RS structures are presented in Fig.\ref{fig8}.

\begin{figure}
\includegraphics[angle=270,width=3.5in]{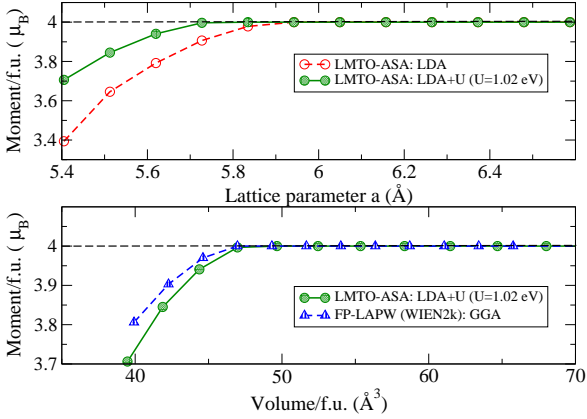}
\caption[]{(Color online) Magnetic moment per formula unit (f.u.) versus lattice parameter for RS CrTe calculated by using LMTO-ASA method and LDA
and LDA+U (U=1.02 eV, J=0) schemes (upper panel). Enhanced correlation effects incorporated via the LDA+U method brings the results in close agreement with those obtained via FP-LAPW-GGA method (lower panel). }
\label{fig9}
\end{figure}

Magnetism is strongly dependent on correlation effects. Even in the metallic states, where magnetism is believed to be described reasonably well
by density functional methods and LDA, different treatment of the exchange-correlation effects lead to different values of magnetic moments. We  thus compare the magnetic moments of our LMTO-ASA calculations for LDA and LDA+U\cite{Anisimov1997}. The enhanced correlation effects in the LDA+U calculations (using U=0.075 Ry, J=0) increase the values magnetic moments and bring the results in general close agreement with those
obtained via using GGA in the WIE2$k$ package. This is shown in Fig.\ref{fig9}. The agreement between GGA and LDA+U results is due to the fact that the GGA goes beyond the LDA by including electron correlations in some way. 

\section{Exchange interaction and Curie temperature}
\label{sec:exchange}
 Exchange interactions are usually calculated by mapping \cite{Sandratskii,Pajda2001} the system energy onto a classical Heisenberg model:
\begin{equation}\label{e1}
H_\mathrm{eff} = - \sum_{i,j} \, J_{ij} \,
{\bf e}_{i} \cdot {\bf e}_{j} \ ,
\end{equation}
where $i,j$ are site indices, ${\bf e}_{i}$ is the unit vector
pointing along the direction of the local magnetic moment at
site $i$, and $J_{ij}$ is the exchange interaction between the moments at
sites $i$ and $j$.   In this work, we use the method of Liechtenstein \etal\cite{Liechtenstein84,Liechtenstein87,Liechtenstein85,Gubanov92}, based on multiple-scattering formalism,
which was later implemented for random  magnetic systems by Turek \etal, using CPA and the TB-LMTO method\cite{Turek2006}. For a discussion of
the various methods used in the mapping procedure the reader may consult our previous publication\cite{Bose2010}.
The exchange integral in Eq.(\ref{e1}) is given by
\begin{equation}\label{eq-Jij}
 J_{ij} = \frac{1}{4\pi} \lim_{\epsilon\rightarrow 0^+} Im \int tr_L \left[\Delta_i(z)g_{ij}^{\uparrow}(z)
\Delta_j(z)g_{ji}^{\downarrow}
\right] dz \; ,
\end{equation}
where $z=E+i\epsilon$ represents the complex energy variable, $L=(l,m)$, and $\Delta_i(z)= P_i^{\uparrow}
(z)-P_i^{\downarrow}(z)$, representing the difference in the potential
functions for the up and down spin electrons at site '$i$'. In the present work 
$g_{ij}^{\sigma}(z) (\sigma=\uparrow,\downarrow)$ 
represents the matrix elements of the Green's function of the medium for the up and down spin electrons. 
The negative sign in Eq.(\ref{e1}) implies that positive and negative values of $J_{ij}$  
are to be interpreted as representing ferromagnetic and antiferromagnetic interactions, respectively.
\begin{figure}
\includegraphics[angle=270,width=3.5in]{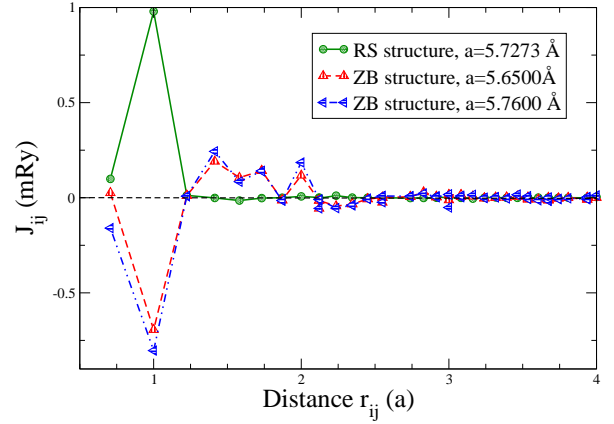}
\caption[]{(Color online) Exchange interaction between the Cr atoms in the ZB and RS structures as a function of the lattice parameter. The two lattice parameters for the ZB structure chosen are above and below the equilibrium lattice parameter 5.727 \AA$\;$ for the RS structure, obtained in the FP-LAPW (WIEN2k) calculation.}
\label{fig10}
\end{figure}

In the mean-field approximation (MFA), the Curie temperature $T_c$, based on the exchange interaction between  the strong moments only, i.e. the
Cr atoms,  can be written as. 
\begin{equation}\label{e2}
k_{B} \, T_{c}^{\rm MFA} = 
\frac{2}{3}\sum_{i \ne  {\rm o}}  J^{\rm Cr,Cr}_{\rm oi} \; ,
\end{equation}
where the sum extends over all the neighboring shells centered around the Cr-site 'o'.
An improved description of finite-temperature magnetism
is provided by the random phase approximation (RPA)\cite{Prange}, with $T_{c}$ given by 
\begin{equation}\label{e3}
(k_{B} \, T_{c}^{\rm RPA})^{-1} = \frac{3}{2} \frac{1}{N} \;
\sum_{\bf q} \, [ J^{\rm Cr,Cr}({\bf 0})-J^{\rm Cr,Cr}({\bf q}) ]^{-1} 
\; .
\end{equation}
Here $N$ denotes 
the order of the translational group applied
and $J^{\rm Cr,Cr} (\bf q)$ is the lattice Fourier 
transform of the real-space exchange integrals $J^{\rm Cr,Cr}_{ij}$.
 $T_{c}^{\rm RPA}$ can be shown to be smaller than
$T_{c}^{\rm MFA}$ \cite{Pajda2001}, and has been found to be in close agreement with results given by
Monte-Carlo simulations \cite{Rusz2006}. Unless otherwise stated, the $T_c$ estimates given in this paper are based on RPA, as given by
Eq.(\ref{e3}).

Sandratskii\cite{Sandratskii} \etal have discussed the case when, in addition to
the interaction between the strong moments, there is one secondary, but much weaker, interaction between the strong and one 
induced moment.
In this case, the Curie temperature, calculated under the mean-field approximation (MFA),
is enhanced due to this secondary interaction, irrespective of the sign of the secondary
interaction. In other words, the Curie temperature would be somewhat higher than that calculated by considering
only the interaction between the strong moments. The corresponding results under the random phase 
approximation (RPA) have to be obtained by solving two equations simultaneously. One can assume that the RPA results for the
Curie temperature follow the trends represented by the MFA results, being only somewhat smaller, as observed in the absence of
induced moments. 

In Fig.\ref{fig10} we compare the exchange interactions between the Cr atoms obtained by using Eq.(\ref{eq-Jij}) for the equilibrium lattice 
parameter of 5.727 \AA$\;$(FP-LAPW GGA result) of RS CrTe with those in ZB CrTe for two values of the lattice parameter, one above and the other below the value 5.727 \AA. For the RS structure the interactions are all ferromagnetic, while for the ZB structure they are predominantly antiferromagnetic in this range of the lattice parameter. In our previous publication we have shown that due to these antiferromagnetic interactions,
the Curie temperature in ZB CrTe is drastically reduced for a certain range of the lattice parameter. The ground state for this structure may even
have a more complex magnetic phase than simple ferromagnetic. Thus for lattice parameters around 5.74-5.75 \AA$\;$ RS CrTe is not only guaranteed to
be a half-metallic ferromagnet, but also should have a higher $T_c$ than ZB CrTe of similar volume per atom.

\begin{figure}
\includegraphics[angle=270,width=3.75in]{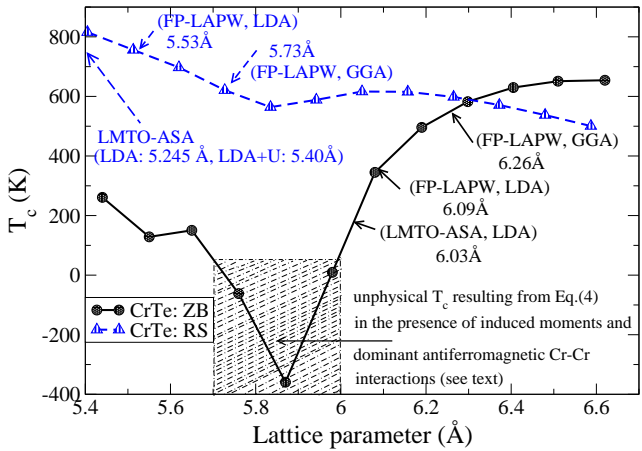}
\caption[]{(Color online) The RPA Curie temperature (Eq.(\ref{e3})) as a function of the lattice parameter for the ZB and RS structures of CrTe, calculated by
considering the Cr-Cr interaction only and by using the FM reference state. The arrows indicate the equilibrium lattice parameter values according to
various calculations. The results for $T_c$ are somewhat affected by the 'induced' moments on the Te atoms as well as the empty spheres used in the LMTO-ASA calculation. For ZB CrTe, this error, combined with strong antiferromagnetic spin fluctuation for the lattice parameters in the range $\sim$5.7-6.0 \AA, result in unphysical negative values of $T_c$ in this range (see text for details). Away from this range $T_c$ values are reliable, although somewhat underestimated.}
\label{fig11}
\end{figure}

\begin{figure}
\includegraphics[angle=270,width=3.75in]{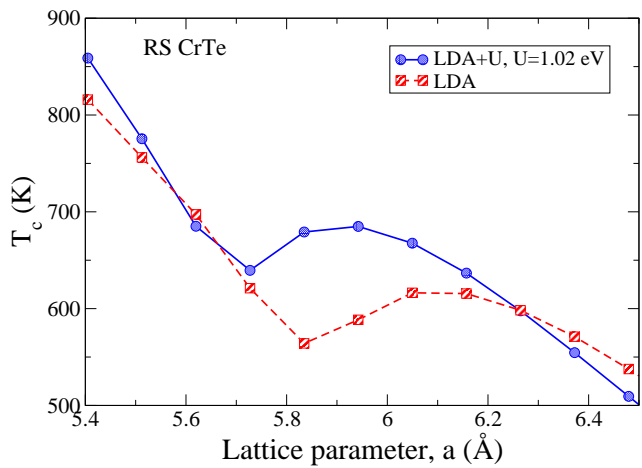}
\caption[]{(Color online) Comparison of the Curie temperature in RS CrTe calculated for the FM reference states in RPA using the LMTO-ASA scheme and LDA and LDA+U methods.}
\label{fig12}
\end{figure}

In Fig.\ref{fig11} we compare the $T_{c}^{\rm RPA}$ values for RS and ZB CrTe obtained by using LMTO-ASA and LDA. The reference state used for the calculation is the FM state. For the ZB phase, the use of Eq.(4) neglecting all the induced moments, results in unphysical negative $T_c$ in the
lattice parameter range $\sim$5.7-6.0 \AA$\;$ because of dominant antiferromagnetic Cr-Cr exchange at the nearest and next nearest neighbor sites.
The ZB phase has been discussed in detail in our previous publication\cite{Bose2010}, where we showed that in spite of these nearest and
next nearest neighbor AFM interactions, the FM state is lower in energy than the AFM[001] and AFM[111] states. The cumulative effect of the FM interactions involving the more distant pairs of Cr atoms, as well as that due to the induced moments, may stabilize the FM phase. Either the cumulative effects of distant pair interactions make the system FM or the magnetic state is other than AFM[001] and AFM[111], possibly complex. If the magnetic state is indeed FM, then according to the prescription of Sandratskii\cite{Sandratskii} \etal $T_c$ should be enhanced with respect to the value given by Eq.(4) and may be positive but small. This explains why Sreenivasan \etal\cite{Sreenivasan2008} and Bi \etal\cite{Bi2008} observe
$T_c$s only as high as 100 K in their thin film sample of ZB CrTe. For RS CrTe, on the other hand, all dominant interactions are FM, and as a result
the $T_c$ calculated by using Eq.(4) and based on only Cr-Cr interactions give positive values. The actual $T_c$s are most probably higher than what is shown in Fig.\ref{fig11}. The effect of enhanced correlation for RS CrTe incorporated via LDA+U method, using U=0.075 Ry, and J=0, is shown in Fig.\ref{fig12}. Within the uncertainties of our results 
there seems to be no appreciable difference between the predictions of LDA and LDA+U.

\begin{figure}
\includegraphics[angle=270,width=4.0in]{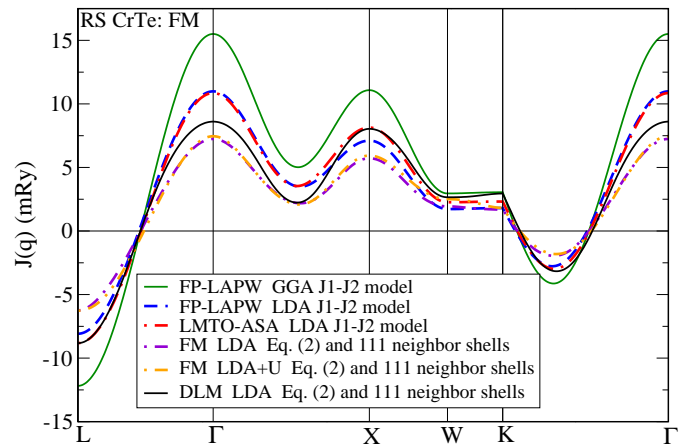}
\caption[]{(Color online) Comparison of the lattice Fourier transforms of the Cr-Cr exchange interactions in RS CrTe for the FP-LAPW-GGA equilibrium lattice parameter of 5.727 \AA$\;$  and  the various cases discussed in the text.
}
\label{fig13}
\end{figure}

It would transpire from Fig.\ref{fig10} that a J$_1$-J$_2$ model may suffice to describe the magnetism of the RS phase. However, for quantitative accuracy of $T_c$ the more distant interactions, although individually small compared to the first two, cannot be neglected, i.e. their cumulative effect is non-negligible. Note that for the ZB phase a J$_1$-J$_2$ model would clearly be inadequate. For the RS phase, if one is prepared to neglect interactions beyond the next nearest neighbor pairs, then the nearest neighbor interaction J$_1$ and the next nearest interaction J$_2$
can be obtained from the energies of the FM and AFM[001] and AFM[111] states\cite{Bruno}. We have calculated J$_1$ and J$_2$ using this
procedure for the FP-LAPW-GGA equilibrium lattice parameter of 5.727 \AA, and the results are: J$_1$=3.76 meV, J$_2$=27.63 meV (FP-LAPW-GGA), J$_1$=3.29 meV,  J$_2$=18.32 meV (FP-LAPW-LDA),
J$_1$=4.61 meV, J$_2$=20.0 meV (LMTO-ASA-LDA). To compare these results with those given by the procedure based on the multiple scattering 
formalism and Eq.(\ref{eq-Jij}) we need to consider that all the distant neighbor interactions between the Cr atoms (and the interactions involving
the induced moments) get mapped effectively on to just two interactions in the J$_1$-J$_2$ model. This is
tantamount to reproducing the band structure of a long-ranged Hamiltonian via a TB-model with only nearest and next-nearest neighbor interactions. Success is possible, but not always guaranteed (unless further interactions are included).
A meaningful way to compare the two approaches is to compute
the dispersion $J({\bf q})= \sum_{{\bf q}}J^{\rm Cr,Cr}_{0{\bf R}}\exp\left(i{\bf q}\cdot{\bf R}\right)$. In Fig.\ref{fig13} we show these 
dispersion curves for RS CrTe along the various symmetry directions and all the different cases considered. The  values of the exchange interactions quoted above for the J$_1$-J$_2$ model would put the $T_c$ for the lattice parameter 5.727 \AA$\;$ around $\sim$1000 K. One reason that the magnitudes of the $J({\bf q})$ values based on Eq.(\ref{eq-Jij}) come out smaller than those given by the J$_1$-J$_2$ model is that in computing $J({\bf q})$ from Eq.(\ref{eq-Jij}) we have
neglected all interactions involving the induced moments.

\begin{figure}
\includegraphics[angle=270,width=4.0in]{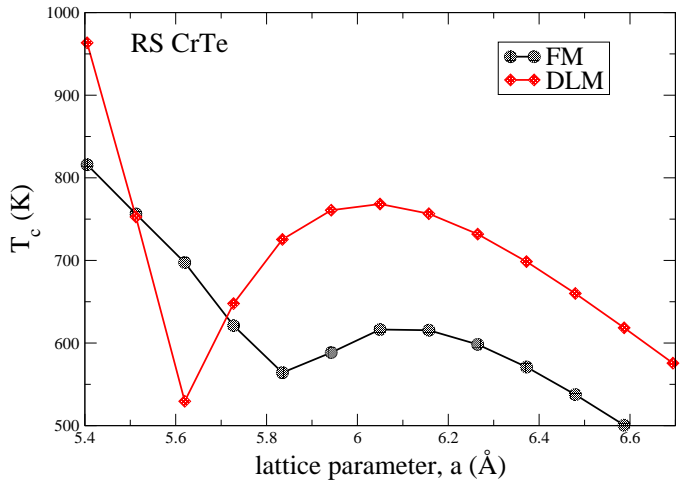}
\caption[]{(Color online) Variation $T_c$ as a function of lattice parameter for FM and DLM reference states in RS CrTe
}
\label{fig14}
\end{figure}

In our previous publication\cite{Bose2010} we have argued that estimates of $T_c$ from above the transition can be obtained by considering the reference state for the exchange calculation in the method of Liechtenstein \etal\cite{Liechtenstein84,Liechtenstein87,Liechtenstein85,Gubanov92} to be the paramagnetic state described  by the disordered local moments (DLM)\cite{Heine1,Heine2,Hasegawa,Pettifor,Staunton,Pindor} model. In this description magnetic moments appear only on the magnetic, i.e. Cr, atoms,
but are oppositely directed for half of the randomly chosen atoms, giving a zero moment in total. The randomness is treated using the CPA\cite{Kudrnovsky1990,Turek97}. $T_c$ given by the DLM reference states are usually higher than their FM counterparts. The DLM estimates of $T_c$ are free from errors originating from the neglect of the exchange interactions involving induced moments in Eqs.(\ref{e2}) and (\ref{e3}). For this reason we have computed the $T_c$ using the DLM model as well and compare these with the $T_c$
obtained for the FM reference states in Fig.\ref{fig14}. As expected, the DLM values are usually higher for most of the lattice parameters. Since 
the ground state of RS CrTe is known to be FM for the entire range of the lattice parameters studied, we expect the FM results to be valid, although
somewhat underestimated due to the neglect of the induced moments. Thus the theoretical estimates of $T_c$ should lie between those given by the FM and DLM results. Experimental results on Cr-based chalcogenides and pnictides show that these estimates are on the high side. One should note, however, that second order phase transition in a bulk material is different from that in thin films grown
on substrates and $T_c$s may differ even though the two transitions belong to the same universality class. A point to note is that for larger volumes per atom the DLM energy is lower than the NM energy, with a Cr-moment that is slightly less than that for the FM state of the same volume. Dominating $J_{ij}$s for the DLM reference states at these volumes are AFM-like. For smaller volumes the Cr-moment disappears, indicating strong hybridization between the orbitals on Cr- and Te-sublattices. 

\section{Summary of results}

We have provided a comparison of the magnetic properties of CrTe in three different structures, NA, RS and ZB, at their respective equilibrium volumes. A detailed comparison of the half-metallic properties of the ZB and RS phases is provided, as thin films grown on most substrates are likely to have either of these two structures. The RS phase is more stable than the ZB phase, in that the lowest energy state in the RS phase is about 0.25 eV/atom lower than that for the ZB phase. For some range of the lattice parameter, namely between 5.6-6.1 \AA$\;$, in the ZB phase there are significant antiferromagnetic interactions between the Cr atoms. The RS phase appears to be free from such strong antiferromagnetic 
interactions. Ferromagnetism appears to be more robust in the RS phase than in the ZB phase. Increased antiferromagnetic interactions in the ZB phase result from the smaller distance between the Cr and the Te atoms and consequently increased hybridization between the Cr-$3d$ and Te-$5p$ orbitals in the ZB phase than in the RS phase for the same lattice parameter. Increased stability with respect to the ZB phase, stronger ferromagnetism and the promise of a higher $T_c$ should make the growth and measurements of the physical properties of RS CrTe an attractive project for experimentalists in the field.

ACKNOWLEDGMENTS 

This work was supported by a grant from the Natural Sciences and Engineering Research Council of Canada. 
J.K. acknowledges financial support from AV0Z 10100520 and the Grant Agency of the Czech Republic (202/09/0775).
Computational facility for this work was provided by SHARCNET, Canada.

\end{document}